\documentclass{aa}  
\usepackage{graphicx}
\usepackage{graphics}
\usepackage{amssymb}
\usepackage{txfonts}
\usepackage{array}
\usepackage{ulem}
\usepackage{longtable,lscape}
\usepackage{array} 
\usepackage{tabularx}
\usepackage{natbib}
\bibpunct{(}{)}{;}{a}{}{,}   

\begin{document}
\title{Milky  Way  demographics with  the  VVV survey\thanks{Based  on
    observations  taken  within  the  ESO  VISTA  Public  Survey  VVV,
    Programme ID 179.B-2002.}\thanks{The VVV survey data are available
    through           the          ESO           archive          {\tt
      http://www.eso.org/sci/archive.html}}}

\titlerunning{Milky Way demographics with the VVV survey}

\subtitle{I.   The  84-million star  colour-magnitude  diagram of  the
  Galactic bulge}

{   
  \author{R.~K.~Saito$^{1,2,3}$
    \and D.~Minniti$^{1,3,4}$
    \and B.~Dias$^{5}$ 
    \and M.~Hempel$^{1,3}$
    \and M.~Rejkuba$^{6}$
    \and J.~Alonso-Garc\'ia$^{1,3}$
    \and B.~Barbuy$^{5}$
    \and M.~Catelan$^{1,3}$
    \and J.~P.~Emerson$^{7}$
    \and O.~A.~Gonzalez$^{6}$
    \and P.~W.~Lucas$^{8}$
    \and M.~Zoccali$^{1,3}$
}
}

\offprints{R. K. Saito: rsaito@astro.puc.cl} 

\institute{
Departamento de Astronom\'{\i}a y Astrof\'{\i}sica, Pontificia Universidad Cat\'{o}lica de Chile, Vicu\~na Mackenna 4860, Casilla 306, Santiago 22, Chile
\and
Departamento de F\'{i}sica y Astronom\'{i}a, Facultad de Ciencias, Universidad de Valpara\'{i}so, Ave. Gran Breta\~na 1111, Playa Ancha, Casilla 5030, Valpara\'{i}so, Chile
\and
The Milky Way Millennium Nucleus, Av. Vicu\~{n}a Mackenna 4860, 782-0436 Macul, Santiago, Chile
\and
Vatican Observatory, Vatican City State V-00120, Italy
\and
Universidade de S\~ao Paulo, IAG, Rua do Mat\~ao 1226, Cidade Universit\'aria, S\~ao Paulo, 05508-900, Brazil
\and
European Southern  Observatory, Karl-Schwarzschild-Strasse 2, D-85748 Garching, Germany
\and
Astronomy Unit, School of Physics and Astronomy, Queen Mary University of London, Mile End Road, London, E1 4NS, UK
\and
Centre for Astrophysics Research, University of Hertfordshire, College Lane, Hatfield AL10 9AB, UK
          }

\date{Received January 01, 2000; accepted January 01, 2000}

\abstract
{The  Milky Way  (MW) bulge  is a  fundamental Galactic  component for
  understanding the formation and evolution of galaxies, in particular
  our  own.  The  ESO Public  Survey  VISTA Variables  in the  V\'{i}a
  L\'actea is  a deep  near-IR survey mapping  the Galactic  bulge and
  southern plane.   Particularly for the  bulge area, VVV  is covering
  $\sim  315$~deg$^2$.  Data taken  during 2010  and 2011  covered the
  entire bulge area in the $JHK_{\rm s}$ bands.}
{We used VVV data for the whole bulge area as a single and homogeneous
  data  set to  build for  the  first time  a single  colour-magnitude
  diagram (CMD) for the entire Galactic bulge.}
{Photometric data in the $JHK_{\rm  s}$ bands were combined to produce
  a single and  huge data set containing $173,150,467$  sources in the
  three  bands, for  the  $\sim  315$~deg$^2$ covered  by  VVV in  the
  bulge. Selecting only the data  points flagged as stellar, the total
  number of sources is $84,095,284$.}
{We built the largest  colour-magnitude diagrams published up to date,
  containing 173.1+ million sources for all data points, and more than
  84.0 million  sources accounting for  the stellar sources  only. The
  CMD  has a  complex shape,  mostly owing  to the  complexity  of the
  stellar  population  and the  effects  of  extinction and  reddening
  towards  the Galactic  centre. The  red clump  (RC) giants  are seen
  double in magnitude at  $b\sim-8^\circ-10^\circ$, while in the inner
  part  ($b\sim-3^\circ$) they appear  to be  spreading in  colour, or
  even splitting into a secondary peak. Stellar population models show
  the predominance of main-sequence  and giant stars.  The analysis of
  the outermost bulge  area reveals a well-defined sequence  of late K
  and  M dwarfs,  seen at  $(J-K_{\rm s})\sim0.7-0.9$~mag  and $K_{\rm
    s}\gtrsim14$~mag.}
{The interpretation of the  CMD yields important information about the
  MW bulge, showing  the fingerprint of its structure  and content. We
  report  a well-defined red  dwarf sequence  in the  outermost bulge,
  which is important  for the planetary transit searches  of VVV.  The
  double RC in  magnitude seen in the outer bulge  is the signature of
  the X-shaped MW bulge, while the  spreading of the RC in colour, and
  even its  splitting into a  secondary peak, are caused  by reddening
  effects.   The  region  around  the  Galactic centre  is  harder  to
  interpret  because   it  is  strongly  affected   by  reddening  and
  extinction.}


\authorrunning{Saito et al.}
\titlerunning{VVV} 
\maketitle
%

\section{Introduction}

The bulge  of the Milky Way  (MW) is a  fundamental Galactic component
for understanding the formation and  evolution not only of our Galaxy,
but of galaxies  in general.  In the MW the  stars can be individually
resolved down  to faint magnitudes, allowing  a complete understanding
of its structure, age, kinematics, and chemical composition.  However,
because the bulge and plane concentrate  not only most of the stars in
the MW, but also gas and dust, observations of the Galactic centre are
difficult  because  they  are  affected by  crowding  and  extinction.
Optical  surveys such  as MACHO  \citep{1999ApJS..124..171A}  and OGLE
\citep{1993AcA....43...69U}  are highly  affected by  extinction while
past  near-infrared (near-IR)  surveys such  as 2MASS  are  limited to
bright             sources            \citep[$K_{\rm            s}\sim
  14.3$~mag;][]{2006AJ....131.1163S}, and do not allow a complete view
of the bulge populations.

\begin{figure*}
\includegraphics[bb=5cm -3.0cm 16cm 10cm,angle=-90,scale=0.55]{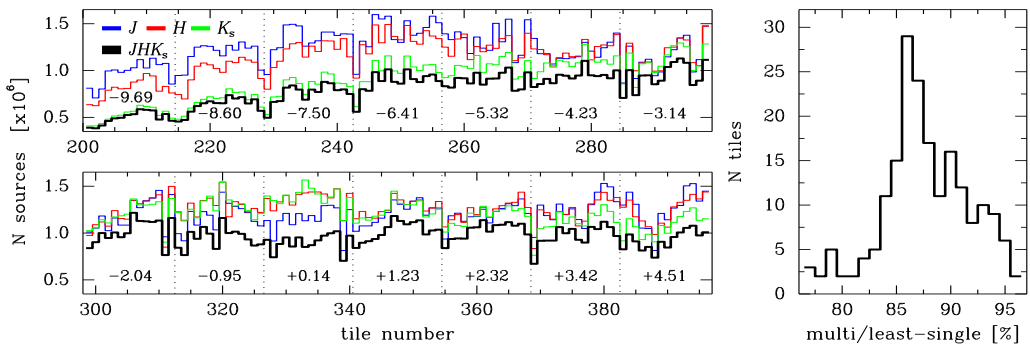}
\caption{Left-hand panels: total number of sources present in the $J$,
  $H$, and $K_{\rm s}$ singleband  and in the $JHK_{\rm s}$ multi-band
  catalogue for each VVV bulge tile.  The tiles are numbered from b201
  to b396,  starting at the bottom-left  corner of the  VVV bulge area
  ($l,b=-10,-10$).   The Galactic  latitude  for the  tile centres  is
  shown in the  figure, while dashed lines mark the  edge of each line
  of  tiles  across  the  bulge.   The  right-hand  panel  presents  a
  histogram  with the  ratio  between  the number  of  sources in  the
  multi-band catalogue  in relation to its  least numerous single-band
  catalogue.}
\label{fig:stat}
\end{figure*}

The  new ESO  Public Survey  VISTA Variables  in the  V\'{i}a L\'actea
(VVV)  is a  deep near-IR  survey mapping  562 square  degrees  in the
southern      plane       and      bulge      of       our      Galaxy
\citep{2010NewA...15..433M}. Particularly  for the bulge  area, VVV is
covering $\sim  315$~deg$^2$. The VVV survey observes  in five near-IR
passbands ($ZYJHK_{\rm  s}$), where extinction effects  are lower than
in  the optical  wavelengths,  and  it is  much  deeper than  previous
near-IR  surveys.  The 5$\sigma$  limiting magnitude  of the  VVV data
using aperture photometry is  $K_{\rm s}\sim18.5$~mag in clean fields,
which allows one to monitor bright  sources such as RR Lyrae and clump
giants  stars along  the whole  Galactic bulge  and  plane, solar-type
stars at the  Galactic centre distances, and even  faint dwarf stars a
few kpc  away. The  VVV data reach  the main-sequence  turn-off (MSTO)
even      in      intermediate-      to     high-extinction      areas
\citep[$E(B-V)\lesssim4.0$,][]{2010NewA...15..433M}.

The  VVV photometry  for  the whole  bulge  area was  combined into  a
single, huge, and  homogeneous data set, allowing us  to build for the
first time  a single colour-magnitude diagram for  the entire Galactic
bulge ($\sim  315$~deg$^2$).  Here we present the  84-million star VVV
colour-magnitude diagram  (CMD) for the  Galactic bulge.  This  is the
largest  CMD  ever published  for  a  large  homogeneous data  set,  a
significant step  forward since the publication of  the 9-million star
CMD  of the  Large Magellanic  Cloud  \citep{2000AJ....119.2194A}.  We
discuss  the  differences  in   the  morphology  caused  by  crowding,
extinction  and sky  brightness. A  forthcoming paper  will  present a
similar  analysis for the  populations across  the VVV  Galactic plane
area.

\begin{figure*}
\includegraphics[bb=1cm -1.6cm 20cm 12cm,angle=-90,scale=0.60]{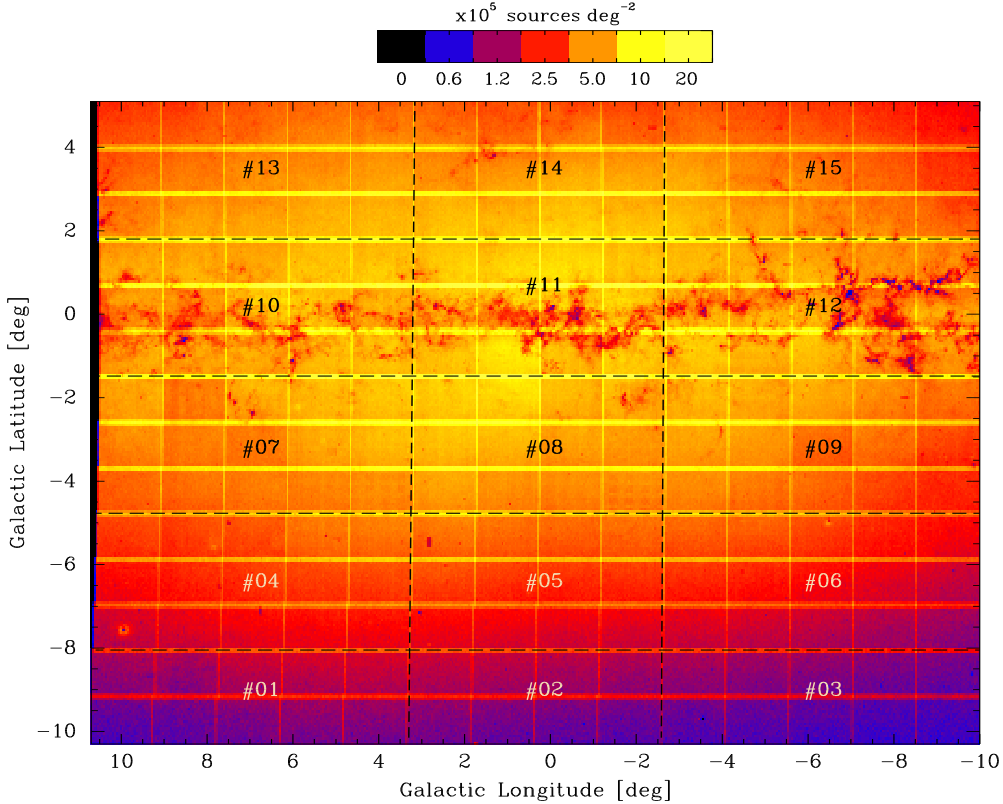}
\caption{Density  plot  in logarithmic  scale  showing  the VVV  bulge
  area. The map was made  using the multi-band $JHK_{\rm s}$ CASU v1.1
  and  v1.2  catalogues  for   point  sources  brighter  than  $K_{\rm
    s}=16.5$~mag (see  text).  Crowded  areas appear in  yellow, while
  less populated regions  as well as highly extincted  areas are shown
  in blue.  The overlapping  regions between the tiles are highlighted
  since the point  sources are accounted for twice.  Dashed lines mark
  the 15  areas selected to build the  colour-magnitude diagrams shown
  in Fig.~\ref{all_15} (see Section  4).  The source density, in units
  of 10$^5$~sources~deg$^{-2}$, is indicated  in the horizontal bar at
  the top. }.
\label{density}
\end{figure*}

\section{Observational data and catalogues}

The VVV is  an ESO Public Survey scanning the MW  bulge and plane with
the 4-m class  VISTA Telescope \citep{2010NewA...15..433M}.  The total
observed  area is  about 562~deg$^2$,  within $-10.0^\circ  \lesssim l
\lesssim +10.5^\circ$ and $-10.3^\circ \lesssim b \lesssim +5.1^\circ$
in the bulge, and within $294.7^\circ \lesssim l \lesssim 350.0^\circ$
and $-2.25^\circ  \lesssim b \lesssim  +2.25^\circ$ in the  plane. The
VVV survey  observes in five passbands, namely  $Z$ (0.87~$\mu$m), $Y$
(1.02~$\mu$m), $J$  (1.25~$\mu$m), $H$ (1.64~$\mu$m),  and $K_{\rm s}$
(2.14~$\mu$m)  bands,  and also  conducts  a  variability campaign  in
$K_{\rm s}-$band  only, with  $\sim$100 pointings spanning  five years
(2010--2014).

Each  unit  of  VISTA  observations  is called  a  (filled)  ``tile'',
consisting of  six individual (unfilled)  pointings (or ``pawprints'')
and covers a  1.64~deg$^2$ field of view.  To fill up  the VVV area, a
total of  348 tiles  are used,  with 196 tiles  covering the  bulge (a
$14\times14$  grid) and  152  for the  Galactic  plane (a  $4\times38$
grid).

\begin{figure*}
\includegraphics[bb=0cm -3.0cm 16cm 15cm,angle=-90,scale=0.55]{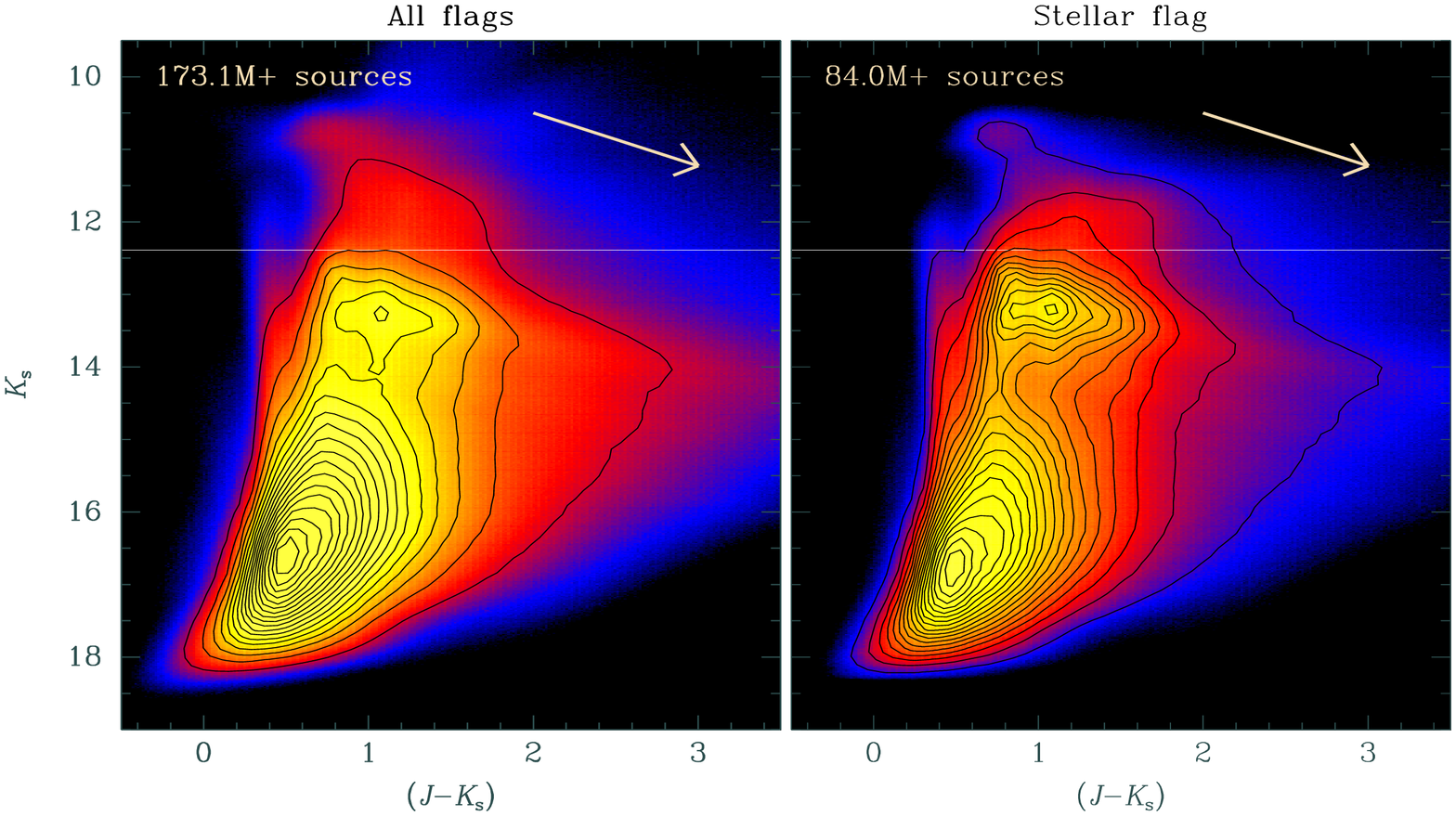}
 \caption{$K_{\rm s}\times(J-K_{\rm  s})$ colour-magnitude diagram for
   the  VVV  bulge  area,  as  shown in  Fig.~\ref{density}.   In  the
   left-hand  panel is  the CMD  for all  point sources  found  in the
   matched  catalogues,  in a  total  of  $173,150,467$ sources.   The
   right-hand panel shows only point sources flagged as ``stellar'' in
   the three  $J$,$H$, and $K_{\rm s}$ catalogues,  according with the
   CASU  photometry.    The  total   number  of  stellar   sources  is
   $84,095,284$.  Contour  lines mark density  levels in steps  of 5\%
   from the maximum density.   The reddening vector associated with an
   extinction of  $E(B-V)=2$, based  on the relative  extinctions of
   the  VISTA  filters,  and assuming  the  \cite{1989ApJ...345..245C}
   extinction law, is also shown in both panels.}
\label{cmd}
\end{figure*}

The schedule  for the first  year (2010) comprised the  observation of
the whole VVV plane and bulge  areas in the five passbands, as well as
five complementary epochs in $K_{s}-$band.  However, delays during the
campaign prevented  the completion  of the planned  observations, with
some of the first year data being observed during 2011.

The  delays mostly  affected  the  $Z$ and  $Y$  observations and  the
variability campaign, while the  $JHK_{\rm s}$ observations were fully
completed during the  first semester of 2011.  From  the 196 pointings
needed  to cover the  total bulge  area, 180  were completed  and made
publicly available  with the VVV Data  Release 1 (DR1)  on August 2011
\citep[a   comprehensive  description   of  this   dataset   is  given
  by][]{2012A&A...537A.107S}.  The  present work is based  on $J$, $H$
and  $K_{\rm  s}$  photometry, from  the  VVV  DR1  and DR2  data  (in
preparation), with complete bulge coverage in the three bands.

Photometric  catalogues  for  the  VVV  images  are  provided  by  the
Cambridge     Astronomical     Survey    Unit     (CASU)\footnote{{\tt
    http://casu.ast.cam.ac.uk/vistasp/}}.  The  catalogues contain the
positions, fluxes, and some shape measurements obtained from different
apertures,  with a  flag  indicating the  most probable  morphological
classification. In particular, we note that ``$-1$'' is used to denote
the best-quality  photometry of stellar objects. Some  other flags are
``$-2$'' (borderline  stellar), ``$0$'' (noise),  ``$+1$' (non-stellar
objects),  ``$-7$''  (sources  containing  bad pixels),  and  ``$-9$''
(saturated sources). Our entire  analysis was performed using the {\tt
  apermag3} aperture fluxes, which are used in the CASU catalogues and
in      the     VISTA     Science      Archive     (VSA)\footnote{{\tt
    http://horus.roe.ac.uk/vsa/index.html}}  as the default  values to
represent the flux.  However, we note that the CASU pipeline starts to
measure   the   positions  and   fluxes   with   the  {\tt   apermag1}
(1\arcsec~diameter) and then successively  increases the aperture by a
factor of $\sqrt  2$ in diameter. For highly crowded  fields as in the
inner Galactic bulge,  using {\tt apermag1} and {\tt  apermag2} can be
more suitable than {\tt apermag3}.

The  VVV  data are  in  the natural  VISTA  Vegamag  system, with  the
photometric  calibration in  $JHK_{\rm s}$  performed using  the VISTA
magnitudes of unsaturated  2MASS stars present in the  images. A final
tile   catalogue   comprises   the   photometry   for   an   area   of
$\sim$1.64~deg$^2$ in the sky, with  the total number of point sources
found  in each  single-band catalogue  ranging from  $408,858$  at the
outermost bulge  region (tile  b202, $K_{\rm s}-$band)  to $1,597,046$
sources  per  tile  at  the  innermost  Galactic  centre  (tile  b246,
$J-$band). The $K_{\rm s}-$band  catalogues are shallower than $J$ and
$H$  at  high  Galactic  latitudes,  while  near  the  Galactic  plane
($|b|\lesssim3$~deg) the three catalogues  present a similar number of
sources  for most  tiles (see  Fig.~\ref{fig:stat}).  On  the Galactic
plane the high extinction mostly  affects the $J-$ and $H-$band, which
renders the $K_{\rm  s}-$band catalogue the one with  the most entries
among the three bands.

The comparison between the  source counts of different tile catalogues
depends strongly, but not  exclusively, on their position with respect
to the  stellar crowding, i.e., the  detection rate for  tiles at high
Galactic latitude will be significantly  higher than for a tile in the
Galactic  plane.  This  effect  becomes even  more  important for  the
aperture-photometry-based  source  catalogues   used  in  this  study.
Artificial  star  experiments  carried   out  on  subsections  of  two
different tiles in the  bulge area \citep[b204 and b314, respectively,
  see][]{2012A&A...537A.107S}   show   that   the   single-band   tile
catalogues  are  80\% complete  down  to  $K_{\rm  s}\sim17.0$ in  the
outermost bulge region  (b204), but in the innermost  bulge (b314) the
same percentage is reached at $K_{\rm s}\sim15.5$.

The single-band CASU catalogues for each tile were matched by VVV team
members, with  most of sources matched within  $\lesssim0\farcs 5$, as
expected   from   the   astrometric   accuracy   of   the   VVV   data
\citep{2012A&A...537A.107S}.   For a couple  of tiles  the ellipticity
sometimes  varies  between  the   filters,  hence  sources  appear  as
high-quality  stars in  one filter  but  are rejected  in another,  or
appear with  a distinct morphological  classification. Therefore, some
sources are  lost during  the matching procedure.  For most  tiles the
multi-band catalogues contain  more than 85\% of of  the total sources
found in the least numerous singleband catalogue used for the matching
(see  Fig.~\ref{fig:stat}).   The   performance  is  maximum  at  high
Galactic  latitude,  with the  multi-band  catalogues  reaching up  to
$\sim$96\% of  successful matches.  On the other  hand, the increasing
source confusion  and extinction  towards the Galactic  centre reduces
the number  of matches in  the innermost region ($|b|<1$~deg),  with a
few tiles  showing $\sim$75-80\% of  the total sources present  in the
least  numerous  singleband  catalogue.   Some larger  ``steps''  seen
between adjacent  tiles in Fig.~\ref{fig:stat} are caused  not only by
differences  in the  quality and  deepness of  data, but  also  by the
obscured regions at lower  Galactic latitudes. There are $173,150,467$
sources in all multi-band catalogues for the 196 bulge tiles.

Figure~\ref{density}  shows  the  VVV  bulge  area as  a  density  map
corresponding to the mosaic  of all multi-band catalogues. To equalize
the  increasing sky  brightness towards  the Galactic  centre  that is
generated  by the  contribution  of the  underlying, unresolved  faint
stars, a  cut in $K_{\rm  s}=16.5$~mag was applied in  the catalogues,
with  a   total  of  $128,660,076$  sources  remaining   in  the  map.
High-density areas appear in  yellow, while less populated regions and
high-extinction areas are shown  in blue.  The tiling pattern produces
overlapping regions  between the tiles.  These are  highlighted in the
map, since the  point sources in these regions  are counted twice. The
overlapping  area  has  been  used  by  the  VVV  team  to  check  the
photometric and astrometric accuracy, but it was not been coadded.

In a few cases the  colour pattern (or source density) changes between
adjacent  tiles, because  some sources  in lower  quality  images were
rejected, as cited above. This effect is mostly seen near the Galactic
centre, where the detection rate is more affected by crowding. Several
tiny dark  spots seen in the  map mark the regions  with no detections
around  bright, saturated  stars.  The  globular cluster  M22  is also
detected at $l,b\sim+9.9,-7.5$. It appears as a region with increasing
density, which  suddenly drops  to a spot  with no detections,  due to
confusion of  the aperture photometry  in detecting sources  in highly
populated fields such as globular cluster cores.

\begin{table}
\caption[]{Effective wavelengths for the  VISTA filter set used in the
  VVV observations  and the relative extinction for  each filter based
  on      the      \cite{1989ApJ...345..245C}      extinction      law
  \citep[from][]{2011rrls.conf..145C}.}
\begin{center}
\label{tab:lambda}
\begin{tabular}{cccc}
\hline \hline 
\noalign{\smallskip}
Filter & ${\rm \lambda}_{\rm eff} (\mu m)$ & $A_X/A_V$  & $A_X/E(B-V)$ \\
\noalign{\smallskip}
\hline
\noalign{\smallskip}
$ J $ & 1.254 & 0.280 & 0.866 \\
$ H $ & 1.646 & 0.184 & 0.567 \\
$ K_{\rm s}$ & 2.149 & 0.118 & 0.364 \\
\noalign{\smallskip}
\hline
\end{tabular}
\end{center}
\end{table}

\begin{figure*}
\includegraphics[bb=-5.2cm 0cm 15cm 19cm,scale=0.53]{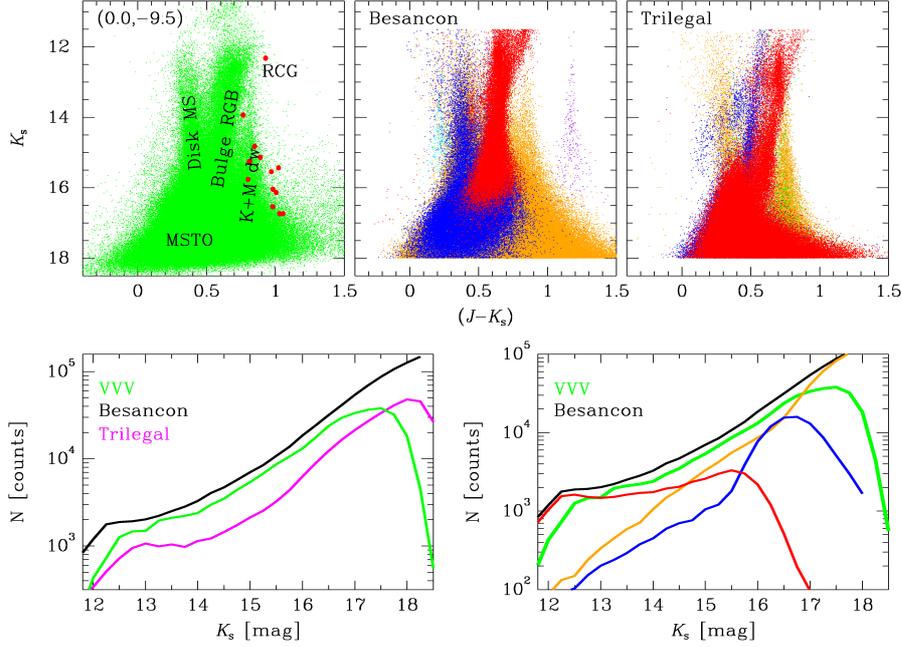}
 \caption{Top-left panel: $K_{\rm s}  \times (J-K_{\rm s})$ CMD for an
   area of  1 sq.  deg.   centred at $(l,b)=(0.0,-9.5)$.  This  is the
   outermost region in the VVV bulge area, on the Galactic minor axis.
   The different  stellar populations are labelled.   Red circles show
   the  colours  of  M-dwarfs  for  a  distance  modulus  of  $(m_{\rm
     Ks}-M_{\rm Ks})=7$~mag (see  text). The top-central and top-right
   panels show  stellar population synthesis models of  the Galaxy for
   the same region, provided by the Besan\c{c}on (centre) and Trilegal
   models (right-hand), including photometric errors and an extinction
   as  described in  the text  (see Section  4).  In  the Besan\c{c}on
   model stars are shown according to luminosity class.  Main-sequence
   stars are depicted in orange,  subgiants in blue and giants are red
   dots.   Supergiants,  bright  giants,  and white  dwarfs  are  also
   present, but there are only a  few (show in light blue, magenta and
   purple, respectively).  The relative  numbers are given in Table 3.
   On  the other  hand,  the Trilegal  model  highlights the  Galactic
   components, with 69.6\% of the  stars belonging the bulge (in red),
   18.3\% are located  in the halo (in blue)  and 10.6\% are thin-disk
   stars (in orange).  A small  fraction of 1.5\% are thick-disk stars
   (green dots).  The bottom-left  panel shows the luminosity function
   for data  and model.  The  bottom-right panel shows  the luminosity
   function  of each luminosity  class in  the Besan\c{c}on  data. The
   colour code is the same as used in the CMDs.}
\label{model}
\end{figure*}

\section{The 84-million star CMD}

We used the  $JHK_{s}$ VVV multiband catalogues of  the complete bulge
as a  single data set  to build the largest  colour-magnitude diagrams
published to date. The left-hand panel of Fig.~\ref{cmd} shows the CMD
containing  $173,150,467$  sources  for  the $\sim  315$~deg$^2$  area
covered by  VVV in the  bulge.  It contains  all sources found  in the
catalogues  and it  is shown  as a  density plot  with  contour curves
marking the same star density levels  in steps of 5\% from the maximum
intensity. The CMD  shows the same data as  in Fig.~\ref{density}, but
also includes the point  sources fainter than $K_{\rm s}=16.5$~mag, in
contrast   with   the  additional   selection   criteria  applied   in
Fig.~\ref{density}.  The  brightest stellar sources  appear at $K_{\rm
  s}\sim9.9$~mag,  while  the  5$\sigma$  limiting  magnitude  depends
strongly on  crowding.  In the  outer bulge the limiting  magnitude is
$K_{\rm s}\sim18.3$~mag,  but at the  Galactic centre the  presence of
underlying,  unresolved faint  stars contributes  to increase  the sky
brightness, which limits the photometry to $K_{\rm s}\sim16.3$~mag.  A
small  fraction of  the sources,  flagged as  ``saturated  stars'' and
``noise'', appears to be populating  the upper and lower limits of the
CMD, respectively.

The right-hand panel  of Fig.~\ref{cmd} shows the CMD  built only with
the data points flagged as ``$-1$'', stellar sources, in the $J$, $H$,
and  $K_{\rm s}$  catalogues, in  a total  of $84,095,284$.  Since the
stellar  flag comprises only  best-photometry data  points, structures
are clearer seen in  the right-hand panel of Fig.~\ref{cmd}. Moreover,
the limiting  magnitude is shallower, since the  faint sources flagged
as ``noise'' are not present. Fig.~\ref{cmd} also shows in both panels
the  reddening vector  associated  with an  extinction of  $E(B-V)=2$,
based  on the  relative extinctions  of  the VISTA  filters using  the
\cite{1989ApJ...345..245C} extinction  law (see Table  1). However, we
notice that  the reddening law  varies significantly in  the innermost
regions of the MW \citep[e.g.,][]{2009ApJ...696.1407N}.

Because the  CMDs were  created by combining  all individual  196 tile
catalogues, they  include the  overlapping regions between  the tiles,
which comprise $\sim7\%$ of the total point sources. Thus, $\sim12.6$M
of the 173.1M  sources are counted twice, which  reduces the number of
unique sources  to $\sim160.4$M.  Similarly, $\sim78.0$M of  the 84.1M
stellar sources are unique.   The overlapping areas are equally spaced
along the bulge  in both $l$ and $b$ directions,  and therefore do not
contribute any bias or special trend in magnitude or colour.

The  following   description,  and  the  next   sections  discuss  the
high-quality CMD that presents the stellar sources.

The  VVV  Bulge  CMD  has  a  complex shape,  mostly  because  of  the
complexity  of the  stellar population,  seen at  different  ranges of
magnitudes and  colours, and the  effects of extinction  and reddening
towards the  Galactic centre. Reddening affects  the bulge populations
at   different   levels,    spanning   from   $E(B-V)\lesssim0.2$   at
$b\sim-10^{\circ}$,  and  up to  $E(B-V)\sim10$  at  the the  Galactic
centre  \citep{2012arXiv1204.4004G}.    Moreover,  the  reddening  law
changes in the innermost bulge regions, as cited above.

In particular, the red giant branch  (RGB) is seen to be broader, with
the main peak  of the red clump (RC) giants  at $(J-K_{\rm s}), K_{\rm
  s}\sim  1.05, 13.20$.   The  RC giants  have  structures ranging  in
colour and  magnitude, with a  secondary peak in colour  at $(J-K_{\rm
  s}),~K_{\rm  s}\sim~0.85,~13.20$,  and  an  elongated  structure  in
magnitude with $\Delta K_{\rm s}  \sim 1$~mag. Those are caused mostly
by the X-shaped bulge of the MW, producing the structure in magnitude,
and by  differential extinction,  resulting in multiple  structures in
colour (see Section 5.1). The colour of the RGB stars is also affected
by changes in the metallicity,  which shows a gradient along the bulge
minor  axis  \citep{2008A&A...486..177Z}.    However,  the  effect  is
minimized in the RC, since the colour of RC giants is less affected by
age and metallicity \citep[e.g.,][]{2002MNRAS.337..332S}.

\begin{figure*}
\includegraphics[bb=-2.5cm 0cm 15cm 14cm,scale=0.68]{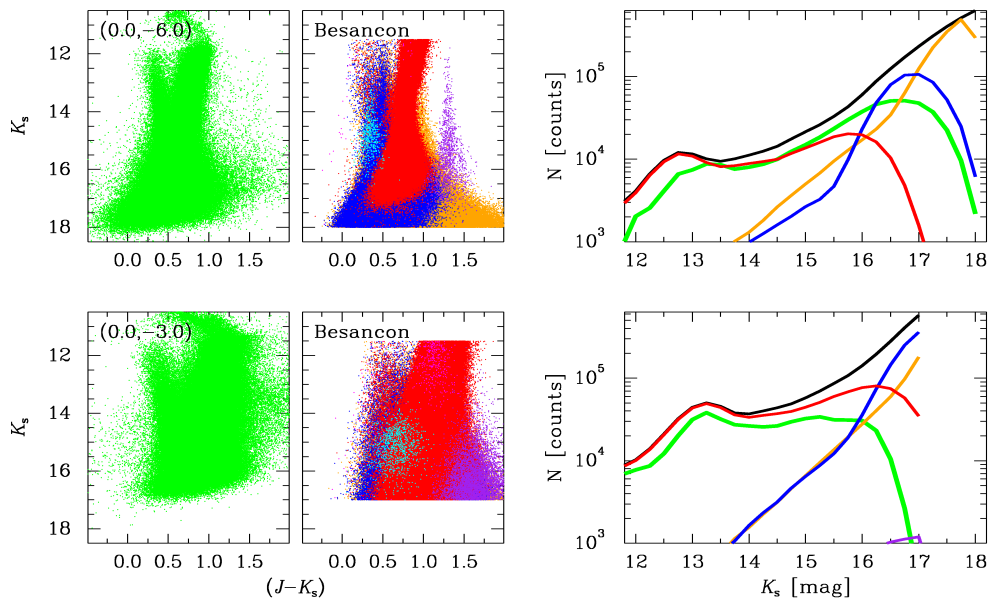}
 \caption{Comparison between  VVV and  synthetic data provided  by the
   Besan\c{c}on model  for the areas at  $(l,b)=(0.0,-6.0)$ (top), and
   $(l,b)=(0.0,-3.0)$  (bottom).   The  notation  is similar  to  that
   presented in Fig.~\ref{model}.}
\label{model2}
\end{figure*}

\section{Comparison with models}

In this section we compare  our data with stellar population synthesis
models      provided       by      the      Besan\c{c}on\footnote{{\tt
    http://model.obs-besancon.fr/}}   \citep{2003A&A...409..523R}  and
the                                              Trilegal\footnote{{\tt
    http://stev.oapd.inaf.it/cgi-bin/trilegal\_1.5}}
\citep{2005A&A...436..895G} codes.

For  the  comparison we  selected  three  regions  of $1^\circ  \times
1^\circ$  along the  bulge.   The  first region  is  at high  Galactic
latitude, centred at $(l,b)=(0.0,-9.5)$.  This is the outermost region
of  the VVV bulge  area, and  is the  least crowded  as well  as least
affected   by    reddening   on    the   bulge   minor    axis,   with
$A_{Ks}=0.023\pm0.010$  (see Table~2). Owing  to these  conditions its
stellar population is well  defined, allowing a better comparison with
models.

The other  two selected regions are centred  at $(l,b)=(0.0,-6.0)$ and
$(l,b)=(0.0,-3.0)$.   These regions are  highly affected  by reddening
and crowding, and are useful for testing the differences in the morphology
of  the CMDs. The  mean values  of reddening  and extinction  for each
region are given in Table 2.

To properly  account for the  differential reddening across the  1 sq.
deg.   regions,  we  divided  each  region  into  16  small  areas  of
$15\arcmin  \times 15\arcmin$  side.  The  mean  $A_{Ks}$, $E(J-K_{\rm
  s})$, and respective  $A_V$ were calculated for each  area using the
reddening     maps    of    \cite{2012arXiv1204.4004G},     and    the
\cite{1989ApJ...345..245C} extinction law.   These values were used as
input  while  querying the  models  for  the  synthetic data  of  each
$15\arcmin \times 15\arcmin$ area.  Finally, the data generated by the
models for  each 16 small areas  were combined into a  single data set
corresponding to the  entire 1 sq.  deg.  This  procedure ensures that
the spatial distribution of sources  across the region and the changes
in the reddening are well-accounted for the synthetic data.

The Trilegal  model is available  in the VISTA $JHK_{\rm  s}$ colours,
while the Besan\c{c}on model uses the $JHK$ system.  Although our data
use  the VISTA  system,  the  systems are  sufficiently  close that  a
comparison is useful for the conclusions.  When synthetic Besan\c{c}on
models  become available  in the  VISTA system,  these should  be used
instead.  The   synthetic  data  were  collected   using  the  default
parameters  presented in  both  models, which  include all  luminosity
classes  belonging the  four  Galactic populations  (thin disk,  thick
disk, halo and bulge) at  the given line of sight, photometric errors,
and extinction as described above.  While the Besan\c{c}on data output
labels  the sources according  to the  luminosity class,  the Trilegal
model  complements  the analysis  because  it  classifies the  sources
according to the Galactic component.

\begin{table}
\caption[]{Mean values for the extinction and reddening for each 1 sq.
  deg.   region  provided   by  \cite{2012arXiv1204.4004G}  using  the
  \cite{1989ApJ...345..245C} extinction law.}
\begin{center}
\label{tab:lambda}
\begin{tabular}{lccc}
\hline \hline 
\noalign{\smallskip}
$(l,b)$ [deg] & $A_{Ks}$  & $E(J-K_{\rm s})$ & $A_V$\\
\hline
\noalign{\smallskip}
($0.0,-9.5$) & $0.023\pm0.010$ & $0.033\pm0.014$ & $0.194\pm0.083$ \\
($0.0,-6.0$) & $0.110\pm0.014$ & $0.160\pm0.020$ & $0.932\pm0.119$ \\
($0.0,-3.0$) & $0.320\pm0.052$ & $0.464\pm0.075$ & $2.710\pm0.438$ \\
\noalign{\smallskip}
\hline
\end{tabular}
\end{center}
\end{table}

The  top-left panel of  Fig.~\ref{model} shows  the $K_{\rm  s} \times
(J-K_{\rm s}$) CMD for the 1 sq.  deg area at $(l,b)=(0.0,-9.5)$, with
labels  indicating  the  distribution  of the  stellar  content.   For
quantitative purposes, the data  include not only the stellar sources,
but all unsaturated sources  presented in the $JHK_{\rm s}$ multi-band
catalogue.   The CMDs built  with Besan\c{c}on  and Trilegal  data are
shown  in the  top-central and  top-right panels  of Fig.~\ref{model},
respectively. Complementary $K_{\rm  s}$ luminosity functions are also
shown  in Fig.~\ref{model}. In  the bottom-left  panel we  present the
luminosity function  of all sources  in each model, compared  with the
VVV  data.   The  luminosity  functions  are quite  similar,  but  the
Besan\c{c}on data are  most numerous than VVV ($+37\%$  sources in the
range $13<K_{\rm s}<17$), while  Trilegal delivers less stars than the
real data ($-49\%$  in the same $K_{\rm s}$  range).  This discrepancy
in total numbers is expected because models based their calibration on
data with different  levels of completeness in the  magnitude range of
the VVV data (e.g., 2MASS, DENIS, OGLE, etc).  However, the difference
in the number  of sources between VVV and  the Besan\c{c}on luminosity
function entirely  agrees with the  results of our  completeness tests
performed  on  the VVV  data  in this  region  (see  Section 2).   The
bottom-right  panel   of  Fig.~\ref{model}  presents   the  luminosity
function  of  the Besan\c{c}on  data  separately  for each  luminosity
class.

We notice that  the double-peaked RC is not  modelled by the synthetic
data  and therefore  does not  allows a  direct comparison.   A recent
paper  by the  Besan\c{c}on team  discuss the  modeling of  the double
clump by introducing  of a flaring bar in the  current version of the
Besan\c{c}on  code,  see  \cite{2012A&A...538A.106R}.  However,  since
they compared the model with 2MASS data, which are not complete at the
fainter magnitude of the second clump, additional modifications in the
model  based  on  deep  near-IR  data are  necessary  (e.g.,  VVV  and
UKIDSS-GPS).

\begin{table}
\caption[]{Relative number  of each luminosity class  in the synthetic
  data provided by  the Besan\c{c}on model.  The range  in $K_{\rm s}$
  was defined according  to the limiting magnitude of  the VVV data in
  each region.}
\begin{center}
\label{tab:lambda}
\begin{tabular}{lccc}
\hline \hline 
\noalign{\smallskip}
\multicolumn{4}{c}{Luminosity classes} \\
\multicolumn{4}{c}{(relative number [\%])} \\
\hline 
\noalign{\smallskip}
\noalign{\smallskip}
 $(l,b)$ [deg] & ($0.0,-9.5$) & ($0.0,-6.0$) & ($0.0,-3.0$)\\
$K_{\rm  s}$ range [mag]  & $11.5-18.0$ & $11.5-18.0$ &  $11.5-17.0$ \\
\hline
\noalign{\smallskip}
Main-sequence & 76.9  & 67.7 & 17.8 \\
Subgiants     & 16.5  & 22.7 & 35.9 \\
Giants        &  6.5  &  9.5 & 45.9 \\
Supergiants   & 0.04  & 0.02 & 0.05 \\
White dwarfs  & 0.05 & 0.06 & 0.3 \\
Bright giants & $<$0.001 & 0.001 & 0.02 \\
\noalign{\smallskip}
\hline
\end{tabular}
\end{center}
\end{table}

The Besan\c{c}on model represents the  VVV data well, showing the disk
and  bulge sequences  at similar  colour and  magnitudes, as  well the
similar scattering  at lower magnitudes  due to the  large photometric
errors.   Main-sequence  stars and  subgiants  are  the most  numerous
sources,  accounting  for  76.9\%  and  16.5\% of  the  total  number,
respectively.   Giants account  for 6.5\%,  but dominate  the  CMD for
$K_{\rm s}<14$~mag.  A small  fraction of luminous supergiants, bright
giants  and white  dwarfs is  also  present.  Table  3 summarises  the
relative  number of each  luminosity class  in the  Besan\c{c}on data.
Finally, the top-right panel of Fig.~\ref{model} shows the CMD for the
Trilegal data.   The sources are labelled according  to their Galactic
components, which are distributed as follows.  The bulge population is
predominant with 69\%  of the total sources.  Halo  stars from farther
out,  behind  the bulge,  account  for  19.6\%,  while the  thin  disk
contributes 10\%.  Thick disk stars make up 1.4\% of the total.

The  comparison  of  our  data  with  the models  for  the  region  at
$(l,b)=(0.0,-9.5)$  allows us  to  conclude that  ({\it\,i}) the  MSTO
highlighted in  the VVV CMD is  composed of bulge,  halo and thin-disk
stars,  with a  majority  of  main-sequence stars,  but  with a  small
fraction of subgiants.   ({\it ii}) The sequence labelled  at the left
side  as  main-sequence  (MS)  stars  is  composed  of  subgiants  and
main-sequence stars from the thin  disk.  ({\it iii}) The bulge RGB at
the right side is composed of giants from the bulge, but also includes
a small fraction of halo  stars behind the bulge.  ({\it iv}) Finally,
a sequence  seen at  the right-most side  is comprised of  late dwarfs
belonging to the thin-disk.

Fig.~\ref{model2} shows the VVV  data compared with the synthetic data
from the Besan\c{c}on model  for the regions at $(l,b)=(0.0,-6.0)$ and
$(l,b)=(0.0,-3.0)$. The increasing  crowding and reddening towards the
Galactic centre introduces degeneracies both in magnitude and colours,
that is why at $b=-3.0$ most  of the sequences are seen overlaid.  The
relative number of each luminosity class also changes, not only due to
intrinsic  changes  in  the  stellar  content, but  also  because  the
increasing  limiting magnitude  of  the VVV  data  that contribute  to
reduce the number of faint sources.  While the number of main-sequence
and subgiants decreases towards  the Galactic centre, the bulge giants
dominate the  CMD at the inner  bulge, increasing by  a factor $\sim8$
from the  CMD at $b=-9.5$ (6.5\% of  the total sources) to  the CMD at
$b=-3.0$ (46\%). A  more detailed comparison with the  models is given
in the next section.

\begin{figure*}
\includegraphics[bb=-1.5cm .5cm 19cm 28cm,scale=0.72]{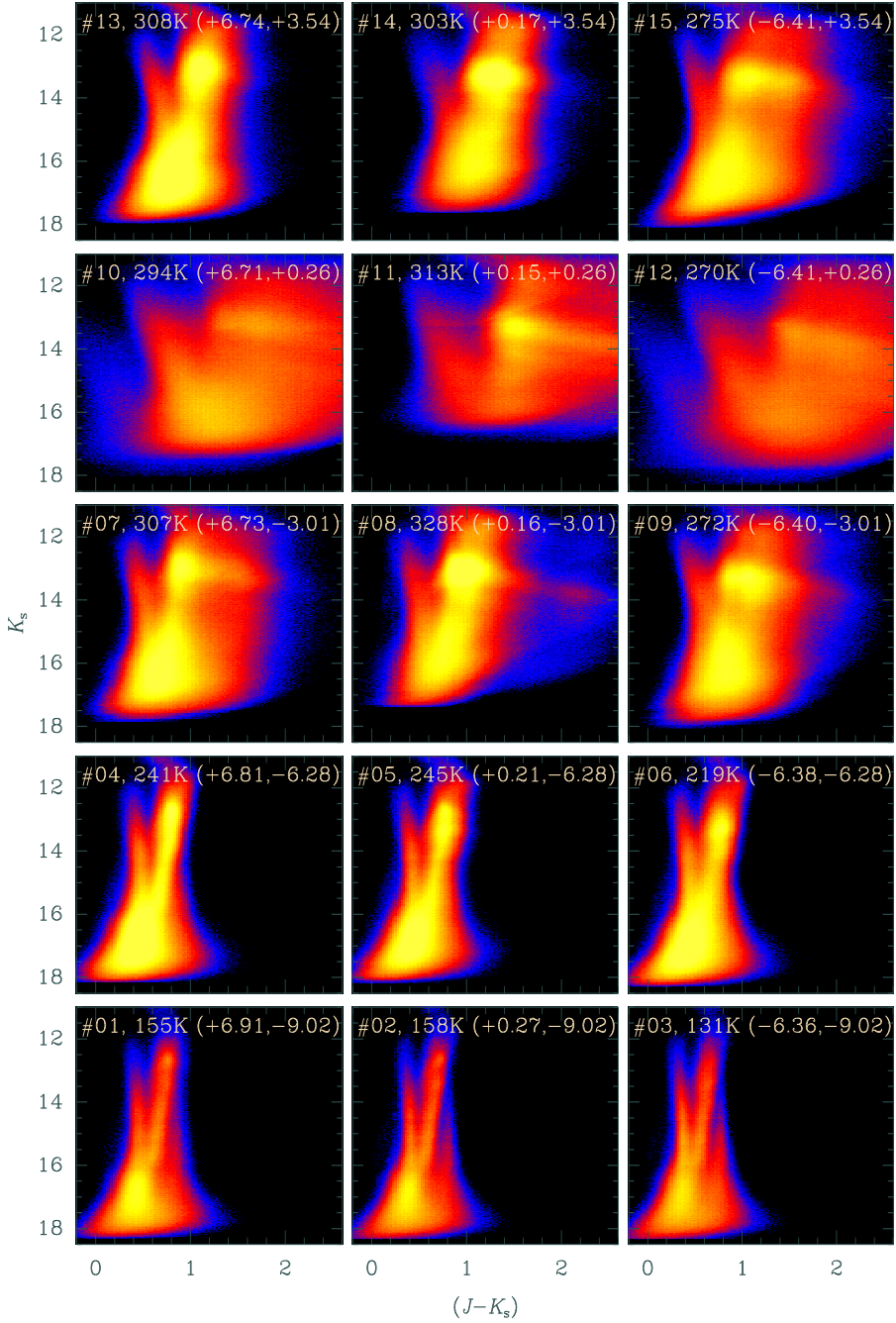}
\caption{CMDs  for 15  different areas  along  the bulge  as shown  in
  Fig.~\ref{density}. The areas were labelled from $\#01$ (bottom-left
  corner) to  $\#15$ (top-right). The mean density  of stellar sources
  (sources~deg$^{-2}$) and  the central coordinates $(l,  b)$ for each
  area are shown in the panels.}
\label{all_15}
\end{figure*}

\section{Slicing the VVV bulge CMD}

To study the bulge CMD in detail we divided the VVV bulge observations
into  15  smaller  areas,  each  comprising  between  8  to  15  tiles
($\sim12-22$~deg$^2$, respectively),  as marked in Fig.~\ref{density}.
This analysis  allows us  to check  by a quick  view the  main changes
caused in the bulge stellar  content by the differential reddening and
extinction towards the Galactic centre.

Fig.~\ref{all_15} shows the 15 CMDs,  with the areas numbered from \#1
to \#15, starting at the bottom-left corner.  The mean stellar density
and  the  central   coordinates  for  each  area  are   shown  in  the
panels. Area \#1 contains not only  MW stars, but also the core of the
Sagittarius dwarf galaxy, in a small proportion of $\sim1:1000$ stars,
according with \cite{2000A&A...357..871C}.

\begin{figure}
\includegraphics[bb=2.5cm 2.0cm 18.cm 15cm,angle=-90,scale=0.99]{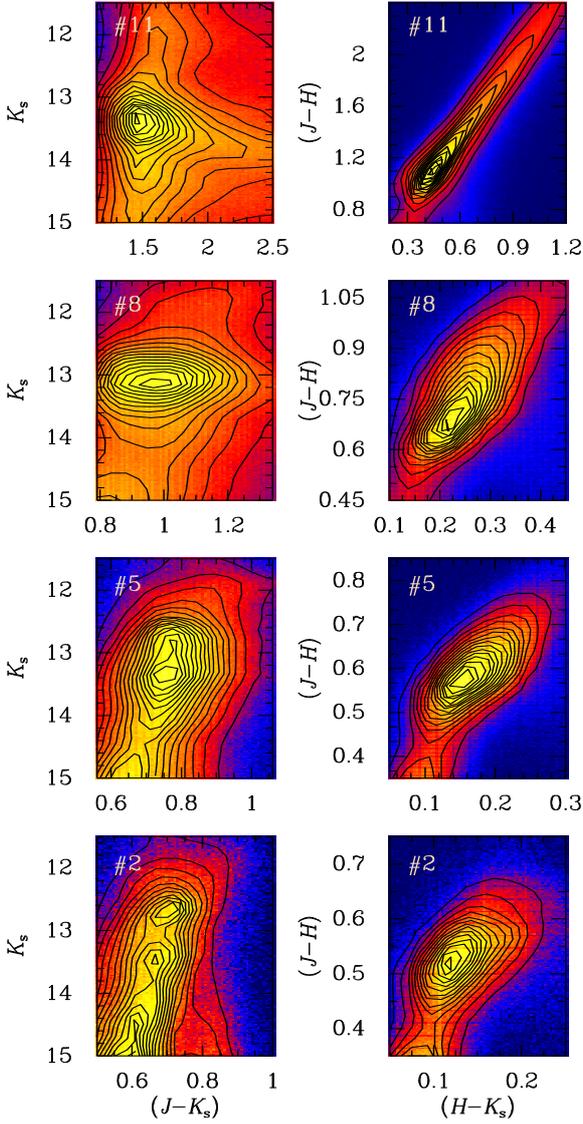}
 \caption{Left-hand  panels:  $K_{\rm s}  \times  (J-K_{\rm s})$  CMDs
   zoomed in  the region  around the Red  Clump giants for  areas \#2,
   \#5, \#8 and  \#11, respectively, from the bottom  to the top.  The
   right-hand panels  show the $(J-H)  \times (H-K_{\rm s})$  CCDs for
   the same areas for all sources brighter than $K_{\rm s}=15$~mag.}
\label{ccds}
\end{figure}

The main differences between the  CMDs occur at the Galactic latitude,
while  differences  in  Galactic  longitude  (which  spans  from  $-10
\lesssim l  \lesssim +10$) at  the same $b$ produces  smaller changes.
Thus, the  central CMD  in each line  of panels  of Fig.~\ref{all_15},
corresponding  to  a given  area  along  the  bulge's minor  axis,  is
representative of its neighbours at higher longitudes. Fig.~\ref{ccds}
highlights the  region around the  red clump giants for  these central
areas on  the minor axis, from  the outermost bulge (Area  \#2) to the
Galactic centre  region (Area \#11).  The figure shows the  $K_{\rm s}
\times  (J-K_{\rm   s})$  CMDs  and   the  $(J-H)\times(H-K_{\rm  s})$
colour-colour diagrams (CCDs) for the given areas.

The RC  of Area \#2 is  found to be double-peaked,  with the brightest
peak  at  $K_{\rm  s}=12.75$~mag  and  the  faintest  one  at  $K_{\rm
  s}=13.45$~mag.  The  peaks appear  with  similar  intensity to  each
other.   The  CCD of  area  \#2 shows  the  RC  as single-peaked,  but
displaced from the  centre of a peanut-shaped structure.   This can be
interpreted as  the sum  of two closer  distributions, as seen  in the
respective CMD.

The second row from the bottom of Fig.~\ref{all_15} shows the CMDs for
areas  \#4, \#5  and  \#6.   These areas  are  moderately affected  by
extinction and  reddening, and the  main structures appear  at similar
colours   and   magnitudes  as   in   the   outermost  region   (areas
$\#1-\#3$). The RC becomes more prominent in relation to the MS and it
is seen extended  in magnitude.  The CMD of  Fig.~\ref{ccds} reveals a
double-peaked  structure, but  now the  farther RC  (the  fainter one,
$K_{\rm  s}=13.40$) is stronger  than the  closest one  (the brighter,
$K_{\rm  s}=12.95$), compared  with  area \#2,  where  the peaks  show
similar intensity. The peaks are closer to each other compared to area
\#2,  with  $\Delta  K_{\rm  s}=0.45$, compared  with  $\Delta  K_{\rm
  s}=0.70$  for the  outermost  bulge. The  CCD  of Area  \#5 shows  a
similar behaviour as  area \#2, with a peanut-shaped  structure with a
single maximum.

Areas \#7,  \#8 and  \#9 are affected  by reddening and  extinction at
different levels, and show the  RC spread over a broader colour range.
Because the star density increases, the unresolved stars contribute to
increase  the background  brightness  limit, with  the faintest  point
sources $\sim0.5$~mag  brighter around the minor axis  (area \#8). The
red giant branch is seen single  in area \#7, with the RC also single,
but spreading  along the  reddening vector.  In  area \#8 the  RC, and
even the RGB seems to be double in colour, but in a closer separation.
This  separation  becomes  larger  in  area \#9.   The  inspection  of
Fig.~\ref{ccds} shows  that the RC main  peak for area  \#8 appears at
$(J-K_{\rm   s})\sim1.05$  over   an   elongated  structure   reaching
$(J-K_{\rm   s})\sim0.90$,   but   at   similar   magnitude   ($K_{\rm
  s})\sim13.15$~mag).  Its CCD shows  similar behaviour, with a single
peak over an elongated peanut structure.

The Galactic centre is sampled in areas \#10, \#11 and \#12. These are
harder to interpret, since they are strongly affected by reddening and
extinction.  The sky brightness is  even higher around the minor axis,
with  the faintest  objects seen  at $K_{\rm  s}\sim16.5$~mag  in area
\#11.   This is the  reason for  the magnitude  colour cut  applied in
Fig.~\ref{density}.  All populations are  seen to be much redder, with
the RC spreading  along the direction of the  reddening vector. The RC
has  a main  peak at  $(J-K_{\rm  s})\sim1.50$, with  a difference  of
$\Delta(J-K_{\rm  s})~\sim~0.8$~mag  in   relation  to  the  outermost
region.

Areas \#13,  \#14 and  \#15 are centred  at $b=+3.54$~deg, and  show a
similar behaviour as seen for areas \#07-09 (centred at $b=-3.01$~deg)
but they are  slightly less affected by extinction.  At the minor axis
(area  \#14) the RC  stars are  more prominent  compared to  the MSTO,
while in the  higher latitudes (areas \#13 and \#15)  both RC and MSTO
have similar strength.

\subsection{The red clump giants}

Owing   to   the    well-defined   absolute   magnitude   and   colour
\citep[e.g.,][]{2000ApJ...539..732A},    and    the    well-understood
dependence          on           age          and          metallicity
\citep[e.g.,][]{2002MNRAS.337..332S}, clump giants have been used with
success as  distance indicators and calibrators  in extinction studies
\citep{1998ApJ...494L.219P,  2002ApJ...573L..51A, 2011ApJ...733L..43M,
  2011A&A...534L..14G,  2012arXiv1204.4004G}.    Our  comparison  with
models presented in  Section 4 demonstrated that clump  giants are the
main population  in the VVV bulge  area, accounting for up  to 46\% of
the total sources at $b=-3.0^\circ$.   Even in the outer bulge, giants
dominate at $K_{\rm s}<14$~mag.

Previous results  have shown  that the double  RC in magnitude  is the
signature        of         the        X-shaped        MW        bulge
\citep[e.g.,][]{2010ApJ...724.1491M,               2010ApJ...721L..28N,
  2010IAUS..265..271Z,  2011AJ....142...76S}.  These  analyses focused
on   intermediate  Galactic  latitudes   ($b\lesssim-6^\circ$),  while
\cite{2011AJ....142...76S} mapped the X-shape structure for the entire
bulge within  $3.5^\circ\leq|b|\leq8.5^\circ$. The RC  splits into two
components for  $|b|\gtrsim5$, with  the separation between  the peaks
increasing  with   the  Galactic  latitude,  in   agreement  with  the
expectation of an X-shape structure seen almost edge-on, with the arms
merging close to  the Galactic centre.  Our data  are complementary to
the  previous results,  and confirm  the persistence  of  the X-shaped
structure for  $b\lesssim-8^\circ$, as seen  in the CMDs of  the outer
bulge (areas \#1-3).  At  area \#2 ($b\sim-9.0^\circ$) the double peak
is  seen very clearly  in the  VVV data,  with the  separation between
peaks reaching $\Delta K_{\rm s} \sim 0.7$~mag (see Fig.~\ref{ccds}).

Combining data  from regions  where the RC  appears with two  peaks at
different separations produces an elongated structure in magnitude for
the final dataset. On the other hand, the differential reddening along
the  bulge,   with  up  to  $E(B-V)\sim10$  at   the  Galactic  centre
\citep{2012arXiv1204.4004G}, causes  the RC stars to  spread along the
reddening vector for  more than $\Delta(J-K_{\rm s})\gtrsim4$~mag. The
combination of  these effects contributes to create  the complex shape
of the RC seen in our CMD for the whole bulge.

\subsection{The dwarf sequence}

The  CMDs of the  outermost bulge  regions (areas  \#1-3) show  at the
right-most side  a well-defined sequence of  relatively faint, $K_{\rm
  s}\gtrsim14$~mag,  and red  objects  $(J-K_{\rm s})\sim0.7-0.9$~mag.
The sequence  merges with  the disk sequence  and the giant  branch at
lower  magnitudes   ($K_{\rm  s}\gtrsim16$~mag),  due   to  the  large
photometric errors in the bottom end of the CMD.  The sequence is seen
to be  stronger towards negative longitudes, becoming  more intense in
the CMD for area \#3.  Interestingly, the sequence disappears at lower
Galactic latitudes, and  it is not visible in any  other region of the
VVV bulge.

The comparison with models presented  in Section 4 shows the red dwarf
sequence as seen in the CMDs for the area at $b=-9.5$~deg.  The models
reveal that the sequence  is composed of main-sequence stars belonging
to  the  thin  disk  (see  Fig.~\ref{model}).  According  to  what  is
observed  in CMDs for  areas \#4-6,  the sequence  is not  detected at
$b=-6.0^\circ$, because of the degeneracies introduced by crowding and
differential reddening, which produce the red dwarf sequence overlayed
with the bulge RGB.

We investigated  the dwarf  sequence in more  detail and  compared the
colours  of M-dwarfs  provided  by \cite{2006MNRAS.367..454H}.   Using
colour transformations  from UKIRT  to the VISTA  system and  the mean
extinction for the  region, M-dwarfs from M1 to  M6 coincide in colour
with  the main  sequence,  while later  objects  are slightly  redder.
Fig.~\ref{model} shows  these objects in the CMD,  assuming a distance
modulus of $(m_{\rm Ks}-M_{\rm Ks})=7$~mag ($d\sim250$~pc).

Interestingly,   the   red  dwarf   sequence   is   seen  at   $K_{\rm
  s}\gtrsim14$~mag,  just beyond  the  limit of  the  2MASS and  DENIS
photometry.  This  can explain  why this was  not noticed  in previous
near-IR   studies.    Deep   CMDs   such   as   those   presented   by
\cite{2003A&A...399..931Z}  should  reveal  the  sequence,  but  their
analysis was focused on fields at $b\sim6.0^\circ$ where the red dwarf
sequence   is   also   absent   in   the  VVV   data,   as   described
above. \cite{2008MNRAS.391..136L} reported a  sequence with late K and
M  dwarfs in the  UKIDSS GPS  data, but  at the  opposite side  of the
Galaxy, at  $l\sim170-175$~deg.  However, the sequence  is not clearly
distinguished from the RGB.  The region at $b=\sim9.0-10$~deg seems to
be ideal to survey these stars  in the Galaxy, because the late dwarfs
are seen in  large numbers.  The quality of  the VVV photometry allows
one to  easily isolate  the sequence using  colour criteria.  K  and M
dwarfs are  particularly important  in the planetary  transit searches
during  the variability  campaign  of VVV,  because  their small  size
allows   a  better   contrast  with   transits  of   low-mass  planets
\citep{2011RMxAC..40..221S}.

\section{The CMD using reddening-free parameters}

The bimodal colour distribution of the red clump can be interpreted as
a reddening  effect generated  by the sampling  of a large  sky region
with enough variation to produce the separation in two components, and
it can be also created  by the observation of two distinct populations
seen at the same line of sight.

In  the  VVV bulge  area  the  RC spreading  in  colour,  or even  its
splitting into  a secondary peak, is  seen across a  large sky region,
area \#9  is the most drastic example.   In this area the  RC and even
the  RGB seems  to be  double in  colour, with  a separation  of about
$\Delta(J-K_{\rm s})\sim0.2$~mag.

To test if the double RC  is caused by differential reddening, we used
reddening-free  parameters   provided  by  \cite{2011rrls.conf..145C},
based  on the  extinction  law of  \cite{1985ApJ...288..618R} for  the
VISTA filters, namely $m_4$ and $c_3$, defined as

\begin{equation}
m_4\equiv K_{\rm s}-1.22(J-H), 
\end{equation}

\begin{equation}
c_3\equiv (J-H)-1.47(H-K_{\rm s}).
\end{equation}

\begin{figure}
\includegraphics[bb=0cm .4cm 16cm 15cm,angle=-90,scale=0.38]{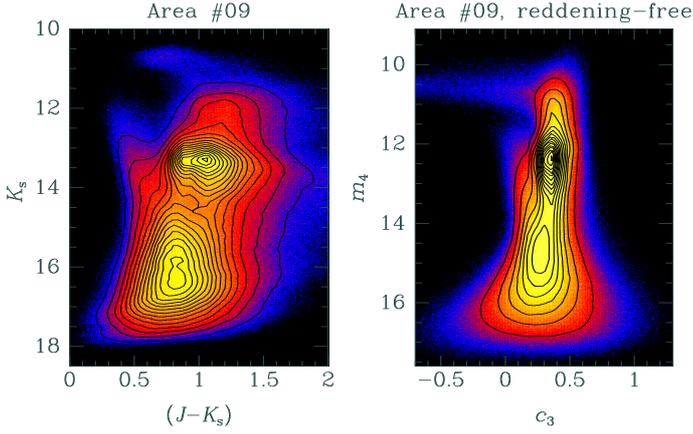}
 \caption{$K_{\rm  s}  \times  (J-K_{\rm  s})$  CMD  for  area  $\#09$
   (left-hand panel)  compared with the CMD  made using reddening-free
   indices (right  panel).  The notation is similar  to that presented
   in Fig.\ref{cmd}.}
\label{red_free}
\end{figure}

Figure ~\ref{red_free} shows the  $K_{\rm s} \times (J-K_{\rm s})$ CMD
for  area  \#9  (left  panel)   compared  with  CMD  built  using  the
reddening-free parameters $m_4 \times c_3$ for the same sources. It is
very clear that the giant sequence  becomes single, and the RC is seen
with a very  sharp single peak.  Not only  the giant sequence changed,
but all structures are seen very concentrated (note the shorter colour
range  in  the  x-axis  of  the  CMD  built  with  the  reddening-free
parameters).

\section{Conclusions}

We presented  the VVV colour-magnitude diagram of  the Galactic bulge,
the largest CMD ever published  for a large homogeneous data set, with
84M+  stars.    The  interpretation   of  the  CMD   yields  important
information  about  the  MW  bulge,  showing the  fingerprint  of  its
structure and content.

Stellar  population synthesis models  fit the  data well,  showing the
predominance  of main-sequence  and giant  stars in  the  outer bulge,
which belong both  to the bulge and halo.   Thin- and thick-disk stars
are also  present in  fewer numbers.   In the inner  bulge the  CMD is
dominated by  bulge giants, which contribute  up to 46\%  of the total
sources at $b\sim-3^\circ$

The  analysis  of the  outermost  bulge  area  reveals a  well-defined
sequence   of   late   K    and   M   dwarfs,   seen   at   $(J-K_{\rm
  s})\sim0.7-0.9$~mag  and $K_{\rm  s}\gtrsim14$~mag. These  stars are
particularly important  in the  planetary transit searches  during the
variability campaign of VVV.

The RC appears to be double in magnitude in the outer bulge region due
to the X-shaped structure of the MW bulge, with the separation between
the peaks  reaching $\Delta  K_{\rm s}=0.70$ at  $b\sim-9^\circ$. This
result  is  complementary with  previous  analyses,  and confirms  the
persistence  of the  X-shaped structure  for  $b\lesssim-8^\circ$.  In
contrast, in  the inner  part ($b\sim-3^\circ$) the  RC appears  to be
spreading in  colour, or even splitting  into a secondary  peak with a
separation of $\Delta(J-K_{\rm s})\gtrsim~0.2$~mag (area \#9). The use
of  reddening-free  parameters confirmed  that  this  last feature  is
caused by reddening effects.

The CMDs of  the Galactic centre are harder  to interpret because they
are strongly affected by reddening and extinction. All populations are
seen to be much redder, and  the RC spreads along the direction of the
reddening vector by $\Delta(J-K_{\rm s})\gtrsim~2.0$~mag.


\begin{acknowledgements}

We  gratefully acknowledge  use of  data  from the  ESO Public  Survey
programme ID 179.B-2002 taken  with the VISTA telescope, data products
from  the Cambridge  Astronomical Survey  Unit, and  funding  from the
FONDAP  Center for Astrophysics  15010003, the  BASAL CATA  Center for
Astrophysics  and Associated  Technologies PFB-06,  the  FONDECYT from
CONICYT, and the Ministry  for the Economy, Development, and Tourism's
Programa  Iniciativa Cient\'{i}fica  Milenio through  grant P07-021-F,
awarded  to The  Milky Way  Millennium Nucleus.   M.R.  and  D.M.  are
grateful for partial support  by the National Science Foundation under
Grant  No.  1066293,  and  the  hospitality of  the  Aspen Center  for
Physics.   B.D.  and  B.B.   acknowledge grants  from CNPq/Brazil  and
FAPESP/Brazil.  J.A.  and M.C. acknowledge support by Proyecto Anillos
ACT-86  and   by  Proyecto   FONDECYT  Regular  No.    1110326.   M.Z.
acknowledges  a fellowship  from  the John  Simon Guggenheim  Memorial
Foundation and support by Proyecto FONDECYT Regular No. 1110393.

\end{acknowledgements}

\end{document}